\documentclass[twocolumn,aps,prb,groupedaddress]{revtex4-1}
  \usepackage{graphicx,wrapfig,amsmath,upgreek,mathtools,amsfonts,physics,multirow,color}
  \usepackage{graphicx}
  \usepackage{color}
  \usepackage{amsmath}
  \usepackage{enumitem}
  \usepackage{amssymb}
  \usepackage{hyperref}

  \newcommand{\be}{\begin{equation}}
  \newcommand{\ee}{\end{equation}}
  
  \newcommand{\bea}{\begin{eqnarray}}
  \newcommand{\eea}{\end{eqnarray}}

  \newcommand{\p}{\partial}

  \newcommand{\lb}{\left[}
  \newcommand{\rb}{\right]}
  \newcommand{\lp}{\left(}
  \newcommand{\rp}{\right)}

  \renewcommand{\vec}[1]{{\boldsymbol #1}}

\begin{document}
  \title{Plasmonic Drag in a Flowing Fermi Liquid} %Doppler Effects in Two-Dimensional Fermi Liquids}
  \author{Haoyang Gao, Zhiyu Dong, Leonid Levitov} %${}^1$}
  \affiliation{Massachusetts Institute of Technology, Cambridge, Massachusetts 02139, USA}

  \begin{abstract}
Collective modes in two-dimensional electron fluids 
show an interesting response to a background carrier flow. 
Surface plasmons propagating on top of a flowing Fermi liquid acquire a non-reciprocal character manifest in a $\pm k$ asymmetry of mode dispersion. The nonreciprocity arises due to Fermi surface polarization by the flow. The flow-induced interactions between quasiparticles make collective modes of the system uniquely sensitive to subtle ``motional'' Fermi-liquid effects. 
The flow-induced  Doppler-type frequency shift of plasmon resonances, arising due to 
electron interactions, can strongly deviate from the classical value. This opens a possibility to directly
probe motional Fermi-liquid effects
in %the ongoing 
plasmonic 
near-field imaging experiments. 
%exhibiting an enhancement due to e-e interactions. This 
%
%%and their relation to 2D plasmons. 
%We identify %and explore 
%several noclassical %Doppler-effect-type contributions 
%motional effects arising due to nonparabolic band dispersion and electron-electron interactions. % between carriers. 
%One is a Doppler-type effect that generates nonreciprocity. Another is a change in % the %collective 
%mode dispersion %symmetric
%that is even in $k$  and does not break time-reversal symmetry,  $\omega(k)=\omega(-k)$. We analyze these % reciprocal and nonreciprocal %Doppler 
%effects for general nonparabolic band dispersion and demonstrate that they experience  nontrivial renormalization due to Fermi-liquid interactions. % in the system. 
%We argue that the %reciprocal Doppler 
%$k$-even effect dominates in the practically relevant 
%long-wavelength regime. % relevant for the ongoing near-field imaging experiments. 
%%interactions on these 
%%In particular, trying to understand how Fermi liquid physics contributes to plasmonic Doppler effect.
  \end{abstract}
%  \date{\today}
  \maketitle
%\tableofcontents

%\section{\bf Motivation}

%\begin{thebibliography}{99}
% \bibitem{dyakonov1993} M. Dyakonov, M. Shur,
% Shallow water analogy for a ballistic field effect transistor: New mechanism of plasma wave generation by dc current, 
% Phys. Rev. Lett. 71 2465 (1993).
%\bibitem{borgnia2015}
%D. S. Borgnia, T. V. Phan, L. S. Levitov,
%Quasi-Relativistic Doppler Effect and Non-Reciprocal Plasmons in Graphene, 	arXiv:1512.09044
%\bibitem{morgado2018} 
%T. A. Morgado, M. G. Silveirinha, 
%Drift-Induced Unidirectional Graphene Plasmons, 
%ACS Photonics, 5, 11, 4253-4258 (2018)
%\bibitem{Correas-Serrano2019}
%D. Correas-Serrano and J. S. Gomez-Diaz, 
%Nonreciprocal and collimated surface plasmons in drift-biased graphene metasurfaces, 
%Phys. Rev. B 100, 081410(R) (2019)
%\bibitem{fizeau} L. D. Landau, E. M. Lifshitz, Electrodynamics of Continuous Media (Pergamon Press, 1960)
%\end{thebibliography}

Plasmonic drag, also known as the plasmonic Doppler effect, is a motional effect that describes %asymmetry 
a change in the dispersion of collective charge oscillations induced by an electric current driven through the system.  
As a simplest case of motional coupling between two different collective flows, the collective oscillations and the DC current,  plasmonic drag is of interest for the quest for new effects due to the electron-electron interactions, and new transport phenomena due to such effects. Graphene plasmonics\cite{wunsch2006,hwang2007,polini2011,koppens2011}, in particular the near-field imaging techniques developed recently\cite{chen2012,fei2012}, provide a platform in which the plasmonic drag effects can be realized and explored. %tools for 

Here we investigate plasmonic drag in a flowing Fermi liquid. The Fermi-liquid interactions are known to be unimportant for plasmons in systems with parabolic  electron band dispersion, where the collective center-of-mass motion of charges can be separated from their relative motion due to the Galilean symmetry\cite{Giuliani1980,Theis1980}.  
However, as we will see, a very different situation occurs for electron systems with a nonparabolic band dispersion such as that of graphene.
%, with the plasmon frequency renormalized by the Fermi-liquid interactions\cite{shtyk_plasmon}. 

In this case, as we will see, the Fermi-liquid interactions do renormalize the Doppler shift. Our analysis, which fully accounts for the interaction effects, predicts the change in the plasmonic frequency in the presence of the flow: %As we will see, 
%%From Eq.\eqref{eq:frequency_model1} we can see 
%The frequencies of the plasmonic mode in the absence and presence of the flow %[$\omega_{0}$ and $\omega_{u}$ respectively] 
%will be shown to be related by
%%\begin{equation}
%%    \omega_{0} =  \sqrt{\frac{v_F p_Fk^2V(k)}{2h^2}}  =\sqrt{\frac{nk^2V(k)}{m^*}}
%%\end{equation}
\begin{equation}\label{eq:model_1_result}
  %  \frac{\omega_{u}}{\omega_{0}} = 1+\frac{k u}{\omega} %\left\lbrace 
    \delta\omega=ku\left(\frac{1}{4}+\frac{3}{4}\frac{G_1+\alpha}{1+F_1} \right)
    +O(u^2)
    . %\right\rbrace
\end{equation}
Here $u$ is the drift velocity, $F_1$ is the $m=1$ harmonic of the Landau interaction and $G_1$ is its radial derivative defined below. The quantity $G_1$, as we will see, %is the Fermi-liquid interaction which 
is uniquely sensitive to the motional effects. The quantity $\alpha$ describes the curvature of the band dispersion, such that $\alpha=0$ for the linearly-dispersing carriers and $\alpha=1$ for parabolic dispersion. This result is valid at relatively weak interactions; a more complicated behavior is found for stronger interactions using a relativistic Landau Fermi-liquid framework\cite{baym1976}
% takes a more complicated form when the interactions become strong. 

% Expansion in $u$ yields the first-order Doppler shift:
%\begin{equation}\label{eq:model_1_result}
%%    \Delta \omega \equiv \omega_{u}-\omega_{0} \approx   \frac{1}{2}\left(\frac{\omega_{u}^2}{\omega_{0}^2}-1\right)\omega_{0}\approx 
%    \delta\omega=ku\left(\frac{1}{4}+\frac{3}{4}\frac{G_1+\alpha}{1+F_1}\right)
%\end{equation}
As a quick sanity check, taking $F_1=G_1=0$ yields the classical Doppler shift $\delta \omega = ku$ when band dispersion is parabolic  [$\alpha=1$]. In contrast, for linear dispersion [$\alpha=0$] Eq.\eqref{eq:model_1_result} predicts %we obtain 
a nonclassical Doppler shift\cite{borgnia2015} $\delta \omega = \frac14 ku$.

%Fermi-liquid renormalization 
This behavior of plasmonic drag displays  an interesting analogy with the seminal results on motional effects in superconducting Fermi liquids\cite{larkin_migdal1963,leggett1965}.
% by Larkin and Migdal, and Leggett\cite{larkin_migdal1963,leggett1965}. 
%The analysis in Refs.\onlinecite{larkin_migdal1963,leggett1965} focused on 
The current-current correlation function, which determines the
response of supercurrent to vector potential, was found to be strongly renormalized by the Fermi-liquid interactions. However, these renormalization effects, while nominally big,  feature a cancellation for systems with parabolic bands.
%expressed through a combination of FL parameters, are nominally big. However, they feature a cancellation for systems with parabolic bands.

%. We immediately see that (a) When non-interacting dispersion is quadratic [$\alpha=1$], Eq.\eqref{eq:model_1_result} restores the conventional Doppler shift $\Delta \omega = ku$. (b) When non-interacting dispersion is linear [$\alpha=0$], Eq.\eqref{eq:model_1_result} restores the relativistic Doppler shift for relativistic plasmon: $\Delta \omega = \frac{1}{4}ku$. 

To emphasize the sensitivity of the Doppler shift to fundamental symmetries of Bloch electrons, such as Galilean symmetry for parabolic bands and Lorentz symmetry for Dirac bands, it
is instructive to make comparison with %Doppler 
%make a comparison to  
light drag in optics. Known as Fizeau drag\cite{fizeau}, it arises due to the speed of light dependence on the velocity of a transparent, moving medium.
%to light drag by a moving medium %in effect in optics 
%describing photon frequency and propagation velocity 
%%energy and momentum change 
%due to motion of the medium. 
For a slowly moving medium, Fizeau drag is %dominated by the 
a $\pm k$-odd effect 
first-order in the medium velocity $u$, %Doppler effect, % known as the Fizeau effect\cite{fizeau},
%for which %is $\pm k$ nonreciprocal %and arises at order 
%such that 
\be\label{eq:fizeau_effect}
%\delta v=u\lp 1-\frac1{n^2}\rp 
%\frac{\delta \omega}{\omega}= \frac{u}{c}
\delta\omega=uk\lp 1-\frac1{\sqrt{n}}\rp 
,
\ee 
where $n$ is the medium refraction index. % and $c$ is the speed of light.  
The reduction of the frequency shift compared to the classical Doppler shift $\delta \omega=%\frac{u}{c}\omega
uk $ is a distinct signature originating from the symmetry of space-time and special relativity. 
%where $u$ and $n$ are the medium velocity and refraction index, and $c$ is the speed of light. %The magnitude of the %nonreciprocal 

For the plasmonic Doppler effect in graphene, our analysis predicts a similar suppression as compared to the classical Doppler effect, a distinct behavior arising due to relativistic carrier dispersion in graphene. At the same time, the effect is 
considerably stronger than the light drag, on the order $\delta \omega/\omega\sim u/v_{\rm p}$, where $v_{\rm p}$ is plasmon velocity. Our analysis also demonstrates that the Doppler shift is further renormalized, and enhanced, by interactions in the flowing Fermi liquid.

The dependence of the Doppler shift on the band curvature $\alpha$ and the electron interactions ($F_1$ and $G_1$) can provide a way to tune the Doppler shift. If band curvature is large and positive, the Doppler shift is enhanced, whereas when curvature is large and negative, the Doppler shift sign is reversed. The interactions $F_1$ and $G_1$ renormalize the Doppler shift and push it away from the free-particle value. Measuring plasmonic Doppler effect can therefore be used to directly probe motional Fermi-liquid effects. Comparison of the effects for different electron band %structures  
dispersion can %help understand 
shed light on subtle aspects of Bloch electron dynamics

In our analysis, we will focus on %the Doppler effect in a 2D 
a two-dimensional Fermi liquid in the collisionless regime
$
\omega\gg\gamma_{\rm ee}
$,
where $\gamma_{\rm ee}$ is the %electron-electron (ee)  
carrier collision rate. In this case, while the effects of collisions are negligible, the effects of ee interactions are not negligible because carriers % in which electrons 
are subject to the short-range Landau interactions in combination with long-range Coulomb interactions. This system is described by the single-particle Hamiltonian
\be\label{eq:H}
H=\epsilon_0(\vec p)+e\phi(\vec x)+\sum_{\vec p'}f(\vec p,\vec p')  n(\vec p',\vec x)
%,\quad
%\Delta n(\vec p',\vec x)=n(\vec p,\vec x)-\bar n
,
\ee
where $\epsilon_0(p)$ is particle dispersion %, the quantity $\bar n$ describes compensating background charge, 
and $\phi(\vec x)$ is the electrostatic potential
\be
\phi(\vec x)=\int d^2x' e \frac{
  n(\vec x')-\bar n
%(n(\vec p,\vec x)-\bar n)
}{|\vec x-\vec x'|}
%,\quad 
 %n(\vec x)=\sum_{\vec p'}n(\vec p,\vec x)
,
\ee
Here $n(\vec x)=\int \frac{d^2p}{(2\pi)^2}n(\vec p,\vec x)$ $\sum_{\vec p'}...$ is the density of distant charges; %denotes $\int \frac{d^2p}{(2\pi)^2}...$, 
the quantity $-\bar n$ denotes compensating background charge due to ions or charge on the gates.  
The last two terms in Eq.\eqref{eq:H} represent the potential energy of a particle due to a change in the distribution of other particles, those far away and those nearby. 
%contribute the $1/|x-x'|$ term, whereas the ones nearby are described by the last %$\sum_{\vec p'}f(\vec p,\vec p') n(\vec p')$ term. 
Distant particles contribute the long-range Coulomb potential %in the term $e\phi(\vec x)$, 
which arises due to a change in the net density of charge at a remote point. The term $\sum_{\vec p'}f(\vec p,\vec p') n(\vec p')$ is the angle-dependent spatially-local interaction of the Landau Fermi-liquid theory. 

We note parenthetically that the apparent singularity at $\vec x'=\vec x$ is an artifact of our decomposition of the potential into a sum of the remote Coulomb part and the local Fermi-liquid part, where `local' and `remote' is defined relative to the Fermi wavelength. While this decomposition is somewhat ambiguous, it will be seen %below 
that the expression above is mathematically sound and well behaved: It  is free from divergences arising at  $\vec x'\approx\vec x$ and provides a correct description in the long-wavelength limit of interest. 

We will write the particle distribution as a sum of the parts describing a steady-state equilibrium in the presence of a flow and a perturbation describing collective charge oscillations:
\begin{align}\label{eq:delta_n}
& n(\vec p,\vec x,t)=n_u(\vec p)+\delta n(\vec p,\vec x,t)
,\quad
\\ \nonumber
& n_u(\vec p)=\frac1{e^{\beta(\epsilon_0(\vec p)+\sum_{\vec p'}f(\vec p,\vec p') n_u(\vec p')
-\vec u\vec p-\mu)}+1}
.
\end{align}
Here the subscript $u$ indicates that the momentum distribution is altered by the flow. 

Since $n_u(\vec p)$ appears under the Fermi function that defines $n_u(\vec p)$, it may seem that the %relation between current and 
dependence of current on the flow velocity $\vec u$ must take a nonclassical form. %However, it turns out that 
%Nevertheless, 
Yet, this dependence takes a completely %classical 
%a very 
conventional form. This can be seen by starting with the expression for current that accounts for a change in velocity due to Fermi-liquid interactions with a $u$-dependent particle distribution:%and transforming it as
%It is described by a selfconsistent Fermi-liquid argument as follows. The latter can be derived by considering a relation between DC current and the quantity $u$:
\begin{align} \nonumber
&\vec j=\sum_{\vec p}e\nabla_{\vec p}\lp \epsilon_0(p) + \sum_{\vec p'}f(\vec p,\vec p')  n_u(\vec p',\vec x) \rp  n_u(\vec p)  
\\
&=\sum_{\vec p}e\vec u n_u(\vec p)  =e\bar n \vec u
.
\label{eq:j}
\end{align}
Here we %calculated current by accounting for the velocity change due to Fermi-liquid interactions with a $u$-dependent particle distribution,  and 
integrated by parts using the identity 
\begin{align} 
&\nabla_{\vec p} \ln (1-n_u(\vec p))
\\ \nonumber
&=\beta \lb \vec v_0(\vec p)-\vec u+\nabla_{\vec p}\sum_{\vec p'}f(\vec p,\vec p')  n_u(\vec p')\rb n_u(\vec p)  
,
\end{align}
where $\vec v_0(\vec p)=\nabla_{\vec p}\epsilon_0(\vec p)$. 

The result in Eq.\eqref{eq:j} identifies the quantity $u$, introduced above as a convenient  parameterization of the flowing carrier distribution, with the drift velocity defined in a conventional way as $j=en v_{\rm d}$. Below we study collective charge oscillations in the presence of the flow and determine the plasmonic Doppler shift by carrying out perturbation theory  in $u$. The relation in Eq.\eqref{eq:j} can then be used to express the Doppler shift through the actual electric current. 

A nonclassical relation that does arise is the one for the Fermi surface displacement induced by the flow. 
Working at small $u$ and assuming a change in particle distribution due to current that happens only near the Fermi level, we can represent the distribution as a displaced Fermi surface
\be\label{eq:p(theta)}
p(\theta)=p_F+\Delta p\cos(\theta)
.
\ee
The amplitude of the displacement $\Delta p$ can be found from the relation defining the Fermi surface, 
\be
\epsilon_0(p)+\sum_{\vec p'}f(\vec p,\vec p')  n_u(\vec p') -\vec u\vec p=\mu
,
\ee 
through rewriting it in terms of the change of the distribution due to the flow
\be
\Delta n(\vec p)=n_u(\vec p)-n_0(\vec p)
.
\ee
As always in the Fermi-liquid theory, it will be convenient to absorb the contribution of a non-moving Fermi sea in the quasiparticle  energy, $\epsilon(p)=\epsilon_0(p)+\sum_{\vec p'}f(\vec p,\vec p')  n_0(\vec p')$. Combining with Eq.\eqref{eq:p(theta)}, we can describe the displaced Fermi surface as 
\begin{align} \nonumber
&0=v_F\Delta p \cos\theta+\sum_{\vec p'}f(\vec p,\vec p')  \Delta n(\vec p')-up_F\cos\theta
\\
&=v_F\Delta p(1+F_1)-up_F\cos\theta
,
\label{eq:displaced_FS}
\end{align} 
where $v_F=d\epsilon(p)/dp$ at $p=p_F$, and we introduced angular harmonics of the Landau interaction defined in the standard way:
\be
F_m=\sum_{p'}e^{-im(\theta_{p'}-\theta_p)}f(\vec p,\vec p')\delta(\epsilon(p)-\mu)
.
\ee
From Eq.\eqref{eq:displaced_FS} we find the relation
\be \label{eq:Delta_p}
\Delta p=\frac{m_* u}{1+F_1}
\ee
where we defined $m_*=p_F/v_F$ the quasiparticle effective mass.

%Here the quantity $\Delta n(\vec p')$ describes the change in the distribution due to the flow
%\be
%\Delta n(\vec p)=
%\ee
%%denotes the last term in Eq.\eqref{eq:H}. 
%
%It is instructive to start with deriving a relation between DC current and the quantity $u$, which differs from the free-particle one by a Fermi-liquid renormalization factor. Working at small $u$ and assuming that the change of particle distribution due to current happens only near the Fermi level, we can write
%\be\label{eq:j}
%\vec j=\sum_{\vec p}\vec v n_0(\vec p)  
%=\sum_{\vec p}\vec v(f*\Delta n-\vec u\vec p)\frac{\p n_0}{\p\epsilon}
%\ee
%Comparing the two terms and passing to angular harmonics on the Fermi surface, we find the relation 
%\be\label{eq:j_u}
%\vec j=\frac1{1+F_1}\sum_{\vec p}\vec v (-\vec u\vec p)\frac{\p n_0}{\p\epsilon}
%=\frac1{1+F_1} n \vec u
%\ee
%where %$n$ is total carrier density. Here 
%we used the relation between carrier density and $D$, the density of states  at $\epsilon=\mu$:
%\be \frac12 p_F v_F D=n.
%\ee 
%Here $F_1$ is the $m=1$ harmonic of the $f(p,p')$ angular dependence times the density of states at $\epsilon=\mu$, defined in the standard way:
%\be
%F_m=\sum_{p'}e^{-im(\theta_{p'}-\theta_p)}f(\vec p,\vec p')\delta(\epsilon(p)-\mu)
%\ee

%\addLL{Do we need to include in Eq.\eqref{eq:j} the $f*\Delta n$ contribution to $\vec v$? Probably not...}

The dynamics of our system is described by classical equations of motion
\be\label{eq:EOM}
\p_t n+\{ H,n\}=0
\ee
where $\{ A,B\}=\nabla_{\vec p}A\nabla_xB-\nabla_xA\nabla_{\vec p}B$ are classical Poisson brackets. We linearize %equations of motion 
the Hamiltonian in the carrier distribution perturbed away from equilibrium as given in Eq.\eqref{eq:delta_n}, arriving at
\begin{align}\label{eq:H_tilde}
& H=\varepsilon(\vec p)+e\delta\phi(\vec x)+\sum_{\vec p'}\tilde{f}(\vec p,\vec p')\delta n(\vec p',\vec x,t)\\
\label{eq:vareps_eps}
&\varepsilon(\vec p) \equiv \epsilon(\vec p)+\sum_{\vec p'} f(\vec p,\vec p') \Delta n(\vec p')
,
%\quad 
%\\ \nonumber
%& \Delta n(\vec p)\equiv n_u(\vec p)-n_0(\vec p)
\end{align}
where $\delta\phi$ is the potential of a distant charge perturbation,  $\delta\phi(\vec x)=\int d^2 x'\frac{e}{|\vec x-\vec x'|}\delta n(\vec x')$. The quantities $\varepsilon(\vec p)$  and $\epsilon(\vec p)$ are the quasiparticle energy in the presence and absence of $u$, respectively; $\tilde{f}(\vec p,\vec p')$ is the Landau function for a shifted Fermi surface. The relation between $\tilde{f}$ and $f$ will be discussed below.
%\be
%\tilde\epsilon(\vec p)=\epsilon(\vec p)+\sum_{\vec p'}f(\vec p,\vec p')\Delta n(\vec p')-\vec u\vec p
%\ee

\begin{figure}
    \centering
    \includegraphics[width=0.5\textwidth]{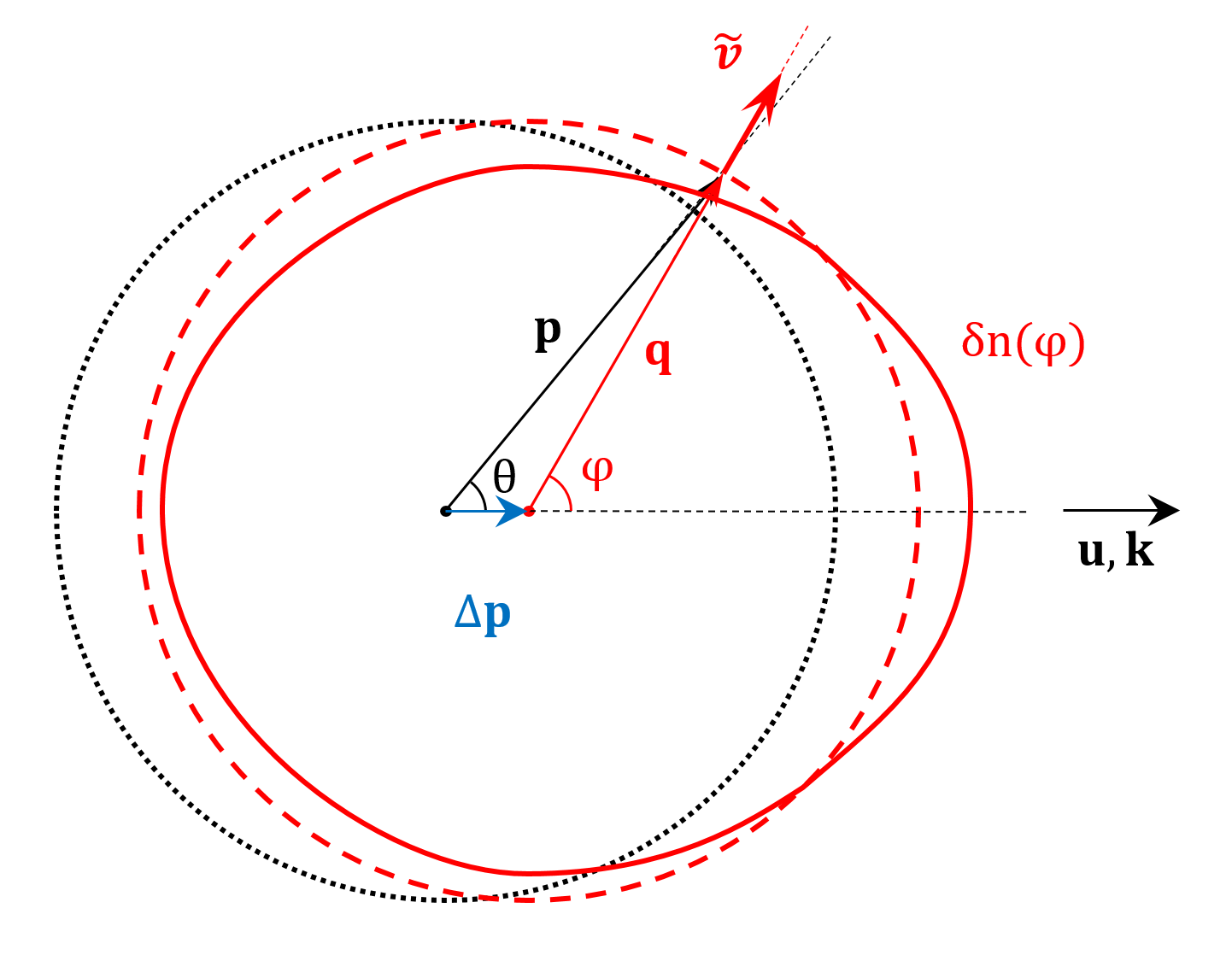}
    \caption{The dashed red and the black dotted lines mark the Fermi surface in the presence and absence of the flow, respectively. Shown are the %The relation between original 
    coordinates $(p,\theta)$ for the Fermi surface at rest, and %shifted 
    the coordinates $(q,\varphi)$ for the Fermi surface describing a flowing Fermi liquid. %Note that 
    Also shown is the vector 
    $\tilde{\vec v}=\nabla_{\vec p}\tilde{\varepsilon}(\vec p)$  normal to the shifted Fermi surface (dashed red line), which is
%    
%    because the former is the gradient of the quantity $\tilde{\varepsilon}(\vec p)$
%    
%    , while the latter is 
    the contour of $\tilde\varepsilon(\vec p) = \varepsilon(\vec p)-\vec p \cdot \vec u$ (see text). 
    %For simplicity, we only study the $\vec k\parallel \vec u$ case.
    } 
    \label{fig:q_varphi}
\end{figure}

To proceed with the analysis, we define $\tilde\varepsilon(\vec p) = \varepsilon(\vec p)-\vec p \cdot \vec u$ %to denote 
the quasiparticle energy in the presence of flow 
%LL
viewed in the comoving frame, then the steady-state distribution describing current flow can be written as
\begin{equation}\label{eq:n_u_tilde}
    n_u(\vec p) = \theta(\mu-\tilde{\varepsilon}(\vec p))
\end{equation}
%LL
%The quasiparticle energy observed in two frames $\tilde{\varepsilon}$ and $\varepsilon$ are related but we do not get into that here.
%by $\tilde{\varepsilon}(\vec p) = \varepsilon(\vec p)-\vec u \codt \vec p$. We note that this is not an exact expression when we consider Lorentz invariance of the graphene electrons. However, at the first order of $u$ this expression is always correct.
%LL
Using this notation and 
%With 
the Hamiltonian in Eq.\eqref{eq:H_tilde}, we linearize equations of motion, Eq.\eqref{eq:EOM}, to obtain
\begin{align}\nonumber
&\p_t \delta n %(\vec p,\vec x,t)
=\lp -e\vec E+\sum_{\vec p'}\tilde{f}(\vec p,\vec p')\nabla_{\vec x} \delta n' %(\vec p',\vec x,t)
\rp \nabla_{\vec p}  n_u(\vec p)
\\ \label{eq:EOM_linearized}
&
-\vec v_{\vec p}\nabla_{\vec x}\delta n %(\vec p,\vec x,t)
%\lp \p_t+\vec v\nabla_{\vec x}\rp \delta n(\vec p,\vec x,t)-e\vec E\nabla_{\vec p}n_0(\vec p)
%+\nabla_{\vec p}n_0(\vec p)\nabla_{\vec x}f*\delta n(\vec p,\vec x,t)=0
\end{align}
where $\delta n$ and $\delta n'$ is a shorthand for $\delta n(\vec p,\vec x,t)$ and $\delta n(\vec p',\vec x,t)$, respectively; 
$\vec E=-\nabla_{\vec x}\delta\phi$ and we defined the velocity in the lab frame $\vec v_{\vec p}\equiv\nabla_{\vec p}\varepsilon(\vec p)$.  %Below we will %use a simple relation between 
%write the quantity $\nabla_{\vec p} n_u(\vec p)$ in terms of $\tilde{\vec v}_{\vec p}$ as
%\be
%\nabla_{\vec p} n_u(\vec p)= -D_{\vec p}\tilde{\vec v}_{\vec p}
%,\quad D_{\vec p}\equiv -\frac{\p n_u}{\p \tilde{\varepsilon}(\vec p)}
%.
%\ee
%This relation follows directly from Eq.\eqref{eq:n_u_tilde}.

To describe collective modes, we consider perturbations of a plane-wave form, $\delta n(\vec p) e^{i\vec k\vec x-i\omega t}$. Writing the field of distant charges as $e\vec E=-i\vec k V(k)\sum_{\vec p'}\delta n(\vec p')$, $V(k)=\frac{2\pi e^2}{k}$, and substituting in Eq.\eqref{eq:EOM_linearized}, gives an integral equation for the collective mode:
\begin{align}\nonumber
&\lp \vec k\cdot \vec v_{\vec p}-\omega\rp \delta n(\vec p)+\vec k\cdot \nabla_{\vec p}n_u(\vec p)\sum_{\vec p'}\tilde{f}(\vec p,\vec p')\delta n(\vec p')
\\ \label{eq:EOM_linearized_w,k}
&=-\vec k \cdot \nabla_{\vec p}n_u(\vec p)V(k)\sum_{\vec p'}\delta n(\vec p')
.
%V(k)\vec k\nabla_{\vec p}n_0(\vec p)\sum_{\vec p'}\delta n(p')
%+\vec k\nabla_{\vec p}n_0(\vec p)F*\delta n(\vec p)=0
\end{align}
Since 
%LL
%Due to the fact that 
the Fermi surface, after being shifted, is still approximately circular (at lowest order in $u/v_F$), we find it convenient to reparameterize all quantities with $\vec q\equiv (q,\varphi)$, denoting the momentum and angle measured from center of the shifted Fermi sea (See Fig.\ref{fig:q_varphi}). 

Using the new coordinate system, all the quantities can be written explicitly. The shifted Fermi sea at zero temperature is simply
\begin{equation}
    n_u(\vec p) = \theta\left(q(\vec p)-p_F\right)
    .
\end{equation}
%LL
%so we find
%\begin{equation}\label{eq:D_p}
%\nabla_{\vec p} n_u(\vec p) = \delta(q-p_F) 
%\end{equation}
%And we re
The perturbed distribution $\delta n(\vec p)$ can be expressed through Fermi surface normal displacement vs. polar angle $\varphi$ on the shifted Fermi surface:
%LL
% the variation of density per unit angle on the shifted Fermi surface $\delta n(\varphi)$, which %by definition 
%is related to $\delta n(\vec p)$ through
\begin{equation}\label{eq:int_varphi}
    \delta n(\vec p) = \frac{h^2}{p_F}\delta(q-p_F) \delta n(\varphi) 
\end{equation}
This relation allows us to convert any integral over $\vec p$ involving $\delta n(\vec p)$ into an integral over $\varphi$:
\begin{align} \nonumber
 &   \int  \frac{d\vec p^2}{h^2} \delta n(\vec p) U(\vec p)  = \int \frac{p_F dq}{h^2} \delta n(\vec p) U(\vec p)  
 \\\label{eq:n_integral}
 &=  \int d\varphi \delta n(\varphi) U(\varphi,q=p_F) 
\end{align}
where $U(\vec p)$ can be an arbitrary function. 

Another %merit 
useful property of the coordinates $(q,\varphi)$ is that the $\varphi$ directly labels the direction of $\tilde{v}$, because being the gradient of $\tilde{\varepsilon}(\vec p)$, the $\tilde{\vec v}$ has to be perpendicular to the shifted circular Fermi surface, which is the contour of $\tilde{\varepsilon}(\vec p)$. This fact will be useful later when we evaluate the velocity component $v^x$.

\end{document}